\documentclass[twocolumn,showpacs,preprintnumbers,prl,floatfix]{revtex4}
\usepackage{epsfig,amssymb,amsmath,hyperref,bm}
\usepackage{epsfig,amssymb}
\usepackage{graphics}
\usepackage{graphicx}
\usepackage{psfrag}
\newcommand\beq{\begin{equation}}
\newcommand\eeq{\end{equation}}
\newcommand\bea{\begin{eqnarray}}
\newcommand\eea{\end{eqnarray}}

\newcommand{\thopp}{t}
\def\myhalf{{\scriptstyle\frac{\scriptstyle 1}{\scriptstyle 2}}}

\begin{document}

\title{Approaching multichannel Kondo physics using correlated bosons:\\
Quantum phases and how to realize them}

\author{Siddhartha Lal}
\email{slal08@illinois.edu}
\author{Sarang Gopalakrishnan}
\email{sgopala2@illinois.edu}
\author{Paul M.~Goldbart}
\email{goldbart@illinois.edu}
\affiliation{Department of Physics and Institute for Condensed Matter Theory,
University of Illinois at Urbana-Champaign,
Urbana, Illinois 61801}

\begin{abstract}
We discuss how multichannel Kondo physics can arise in the setting of a localized level coupled to several bosonic Tomonaga-Luttinger liquid leads.
We propose one physical realization involving ultracold bosonic atoms coupled to an atomic quantum dot, and a second, based on superconducting nanowires coupled to a Cooper-pair box.
The corresponding zero-temperature phase diagram is determined via an interplay between Kondo-type phenomena arising from the dot and the consequences of direct inter-lead hopping, which can suppress the Kondo effect.
We demonstrate that the multichannel Kondo state is stable over a wide range of parameters.
We establish the existence of two nontrivial phase transitions, involving a competition between Kondo screening at the dot and strong correlations either within or between the leads (which respectively promote local number- and phase-pinning).
These transitions coalesce at a self-dual multicritical point.

\end{abstract}

\pacs{72.15.Qm, 71.10.Pm, 73.23.-b}

\maketitle

\noindent
{\it Introduction.}
Magnetic impurities can drastically alter the low-temperature properties of metals, leading to anomalous temperature dependence in, e.g., the heat capacity, resistance, and magnetoresistance.  These properties, collectively known as the Kondo effect~\cite{kondo}, are exhibited when the impurity interacts antiferromagnetically with the conduction electrons of the metal, and are due to the dynamic screening, as $T\rightarrow 0$, of the spins of the individual impurities by a cloud of electrons. The low-energy scattering properties of each screened impurity are those of a bound but spin-polarizable impurity-cloud \textit{singlet}, whose effects on the electron gas can be described using Fermi-liquid theory~\cite{nozieres}. For a ferromagnetic impurity-cloud interaction, on the other hand, the spins of the impurity and its polarization cloud align in one of two degenerate \textit{triplet} ground states. A quantum phase transition separates the ferro- and antiferromagnetic (i.e., Kondo) cases.
\par
A striking generalization of this \lq\lq $1$-channel\rq\rq\ Kondo effect is the multichannel version~\cite{nozblan}, in which each impurity couples \textit{separately} to conduction electrons that propagate in $N(>1)$ channels (e.g., distinguished by orbital angular momentum).  When $N > 2s$ (with $s$ being the spin of the impurity), the conduction electrons \textit{overscreen} the impurity, and the low-energy behavior can no longer be described by Fermi-liquid theory; instead, the thermodynamic properties follow anomalous power-laws governed by a quantum critical point~\cite{afflud}.
In metallic systems, observing the $N$-channel Kondo effect has proven demanding~\cite{coxzawa}.
Although advances in nanoscience have enabled the exploration of Kondo physics in the more controlled setting of 2DEG-based quantum dots~\cite{goldhabergordon} connected to leads (which realize the channels), even here the engineering  of multichannel Kondo phenomena remains a challenge.
This is primarily due to the difficulty in preventing interchannel hybridization, which gives rise at low energies to a single composite channel that screens the impurity via the 1-channel Kondo effect.
For instance, signatures of the two-channel Kondo effect have recently been observed in a quantum-dot-based setting~\cite{potok}, but required fine tuning to prevent hybridization.
\par
In view of these challenges, it is desirable to explore the $N$-channel Kondo effect and its onset in a more readily controllable setting: this can be achieved, e.g., using leads having tunable interparticle correlations.
For the 1-channel Kondo effect the possibility of such control was demonstrated  in Ref.~\cite{furusaki:matveev} in the context of a quantum dot coupled to a Tomonaga-Luttinger liquid (TLL) lead~\cite{giamarchi}.  The strength of the repulsive interactions in the lead (as encoded in the TLL parameter $K$) was found to tune the position of the Kondo-to-ferromagnetic phase transition.
Motivated in part by this result, in this paper we explore the case of $N$ TLL leads coupled to a quantum dot; this case is expected to exhibit richer physics arising from the possibility of not only intra-lead but also inter-lead correlations.
We focus on the case of bosonic leads because for them a wide range of $K$ values can be experimentally accessed with relative ease: e.g., ultracold bosons with short-range repulsions have $K > 1$~\cite{dalibard}, whereas ones with dipolar interactions have $K < 1$~\cite{kollath}.
We suggest two concrete realizations: one uses cold atoms~\cite{paredes,recati}; the other uses superconducting nanowires and a Cooper-pair box~\cite{cpb}.
Systems of ultracold atoms are especially well suited to the study of multichannel Kondo physics because their interactions are highly controllable and tunable~\cite{dalibard}, and extraneous noise can be mitigated.
\par
Our main theoretical results concern the competition between the $N$-channel Kondo effect and interactions in the leads; these interactions either suppress lead-dot tunneling or generate inter-lead phase-locking (which would short-circuit the dot).
Our analysis yield a phase diagram containing four distinct phases (see Fig.~\ref{kondorg}, below) and exhibiting a pair of unusual phase boundaries, which meet at a self-dual multicritical point.
As discussed below, an important advantage of our proposed experimental realization of the $N$-channel Kondo effect is that it is robust against interlead hybridization.
We discuss the conditions for observing the $N$-channel Kondo effect using cold atoms, and suggest that the unusual phase boundaries may be accessed experimentally using dipolar bosons.
\par
\noindent
\textit{Kondo and resonant-level models.}
We consider the general anisotropic Kondo Hamiltonian
$H_K = H_{\mathrm{leads}} + H_{\mathrm{int}}$,
where $H_{\mathrm{leads}} =
\sum\nolimits_k v_F \,k \,c^\dagger_{k\sigma}\, c^{\phantom{\dagger}}_{k\sigma}$
describes a free-electron gas (or a Fermi liquid) and
$H_{\mathrm{int}}$ describes the coupling to the impurity spin, which is located at ${\bf r}=0$:
\[
H_{\mathrm{int}}\! =\! \frac{J_\perp}{2}
\left[S_+ c^\dagger_\downarrow(0) c^{\phantom{\dagger}}_\uparrow(0) + \mathrm{h.c.}\right]
+ \frac{J_z}{4} S_z
\sum\nolimits_\sigma \! \sigma c^\dagger_\sigma(0) c^{\phantom{\dagger}}_\sigma(0),
\]
where $c_\sigma(0)$ annihilates a conduction electron of spin $\sigma$ at the impurity location,
the Pauli matrices ${\bf S}$ act on the impurity spin state, and
$J_z$ and $J_\perp$, respectively, the amplitudes for
the lead-dot Ising and spin-flip processes.
The antiferromagnetic case, $J_{z}> 0$, leads to Kondo screening;
for $J_{z} < 0$ and $|J_\perp| < |J_{z}|$,
the impurity couples ferromagnetically to the conduction electrons.
\par
We shall primarily be concerned with a model that is equivalent to the Kondo model, viz., the interacting resonance-level model (iRLM). This model consists of
a localized level $d$ (i.e., a quantum dot) at the Fermi energy
hybridized with $N$ channels of \textit{spinless} noninteracting conduction electrons $c^{\phantom{\dagger}}_{j{\bf k}}$,
together with a short-ranged repulsion between dot and lead electrons:
$H_{\mathrm{iRLM}}=H_{\mathrm{leads}}+H_{\mathrm{onsite}}+H_{\mathrm{int}}$.
Here, the leads are described by
$H_{\mathrm{leads}}=
\sum_{{\bf k}}
\sum_{n=1}^{N}
\epsilon({\bf k})\,
c^{\dagger}_{n{\bf k}}\,
c^{\phantom{\dagger}}_{n{\bf k}}$,
the on-dot potential by
$H_{\mathrm{onsite}}=B\,d^{\dagger}\,d$,
and the dot-lead couplings by
\def\mybsize{\big}
\beq \label{eq:irlm}
H_{\mathrm{int}}^{n}\!\!=\!\!
V\mybsize(d^{\dagger}c^{\phantom{\dagger}}_{n}(0) + \mathrm{h.c.}\mybsize)+
U\mybsize(d^{\dagger}d-\myhalf\mybsize)
\mybsize(c^{\dagger}_{n}(0)\,c^{\phantom{\dagger}}_{n}(0)-\myhalf\mybsize)
\eeq
where $V\sim J_\perp$ and $U \sim J_{z}+\mathrm{const}$.
The canonical transformation that maps the Kondo model on to the iRLM consists of identifying the \textit{spin\/} density waves of the Kondo model with the \textit{particle\/} density waves of the iRLM.  The analogy is most evident as $U\rightarrow\infty$: the presence of an electron on the dot ensures the absence of electrons near the dot, and vice versa, i.e., an anticorrelated state resembling the Kondo singlet.  The Kondo effect manifests itself in the iRLM via an enhancement, as $T\rightarrow 0$, of number fluctuations on the dot.
\par
We now turn to the iRLM with TLL leads,
governed by the Hamiltonian
$H_{\mathrm{leads}} =
\sum\nolimits_k E(k)\,b^\dagger_k b^{\phantom{\dagger}}_k$,
in which $\{b_k\}$ are bosonic fields that describe free collective phonon modes that disperse linearly~\cite{giamarchi}: $E(k)\sim\vert{k}\vert$.
As discussed in Ref.~\cite{furusaki:matveev}, this version of the iRLM can also be mapped on to the standard Kondo model, but the location of the ferro- to antiferromagnetic transition depends on $K$.
The boundary between the ferro- and anti-ferromagnetic phases occurs at
$U = \hbar v_s (\sqrt{1/2K} - 1)$, in which $v_s$ is the speed of sound in the leads.
\par
In the $N$-channel generalization of the iRLM, each channel couples independently to the impurity via $H_\textrm{int}^{n}$.
By inverting the iRLM-Kondo mapping, one arrives at the $N$-channel Kondo Hamiltonian studied in Refs.~\cite{nozblan,afflud}, in which each of $N$ channels couples independently to the spin.
However, the interchannel particle transfer terms (i.e., hybridization)
$H_\Gamma^{nn^{\prime}}\sim
c^\dagger_{n\alpha}(0)\,
\sigma_{\alpha\beta}\,
c^{\phantom{\dagger}}_{n^{\prime}\beta}(0)$,
which destabilize the $N$-channel Kondo effect, do not arise in the iRLM.
This is because the particle-number (or \lq\lq charge\rq\rq) sector of the leads in the iRLM---the model of physical relevance here---maps on to the \textit{spin} sector of the equivalent Kondo model.
In contrast, the \textit{particle} sector of the leads in the equivalent Kondo model has no physical significance in the iRLM and, accordingly, cannot couple to the dot.
\par
\noindent
\textit{Realizations of the iRLM.}
Our first realization involves ultracold atoms, and extends the ideas of Ref.~\cite{recati}. A star-shaped pattern of $N$ one-dimensional leads meeting at a point can be constructed by passing a laser beam through a phase mask or spatial light modulator (SLM)~\cite{dholakia}.  Such a device is a sheet of glass of spatially varying thickness, which distorts flat wavefronts, giving rise to a prescribed intensity pattern at a ``screen\rq\rq\  some fixed distance away. Algorithms for the construction of appropriate phase patterns are discussed in Refs.~\cite{dholakia, pasienski}.
\begin{figure}[htb]
\begin{center}
\scalebox{0.8}{
\includegraphics{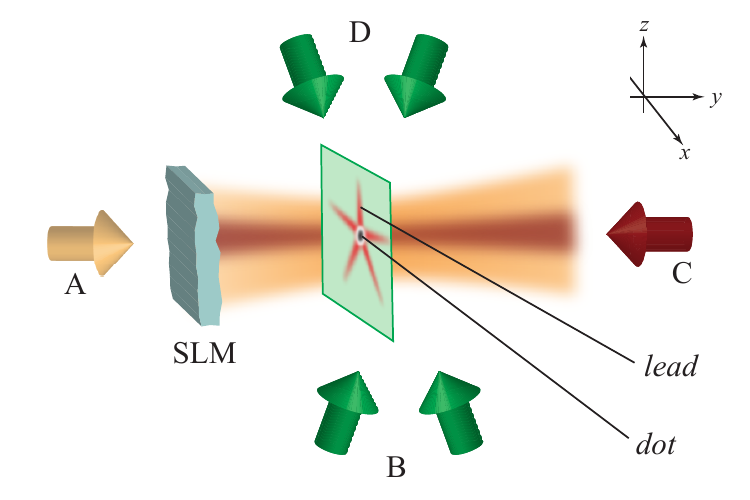}}
\end{center}
\vspace*{-0.6cm}
\caption{\textbf{Candidate cold-atom setup.}
The  spatial light modulator (SLM) distorts the wavefronts of laser~A so as to create a star-shaped pattern at the ``screen\rlap.\rq\rq\thinspace\  Atoms are confined to the screen using two lasers~(B) propagating at a relative angle $\theta \ll \pi$ so as to create an optical lattice of spacing
$L = \lambda/[2\sin (\theta/2)]$. Lasers C and D are used to confine atoms in hyperfine state $b$ at the
``dot\rq\rq\ (dark blue region).}
\label{coldatoms}
\end{figure}
\par
As for the dot itself, it can be realized as follows~\cite{recati}. Suppose that the atoms discussed in the preceding paragraph are in a hyperfine state $a$.  A tightly-confining trap for a different hyperfine state $b$ is now created at the intersection of the leads.  An atom in state $a$ can make a Raman transition to state $b$, and  vice versa; while in $b$, it is confined at the ``dot\rlap.\rq\rq\thinspace\ The Raman transition thus creates a lead-dot hopping amplitude. This setup realizes an iRLM having the following couplings: $K$ is determined by the scattering length $g_{aa}$ for atomic state $a$, $U$ by the $a\leftrightarrow b$ scattering length $g_{ab}$, and $V$ by the amplitude (i.e., effective Rabi frequency) $\Delta$ of the Raman transition. Double occupancy of the dot is prevented by a large repulsive interaction $g_{bb}$ between atoms on the dot (with $g_{bb}\gg g_{aa}, g_{ab}$). All interactions are tunable via a Fesh{\-}bach resonance.  The direct interlead hopping amplitude is governed by the intensity of the laser that traps $a$-state atoms at the intersection of the leads. 
\par
One can realize a similar model in a mesoscopic setting by using superconducting nanowires as the leads, together with a Cooper-pair box~\cite{cpb}---i.e., a superconducting island that holds at most one Cooper pair---as the dot. The normal modes of the leads are plasmon excitations, and the Hamiltonian for the box-lead system has the same form as that for the cold atom system, provided the leads are connected to the dot via Josephson couplings (which determine $V$). The coupling $U$ is determined by the lead-dot Coulomb repulsion.
\par
\noindent
\textit{Analysis of the model.}
In addition to processes that involve the dot, we account for those in which bosons hop directly between leads.
It is useful to write the Hamiltonian for the uncoupled leads as $H_{\mathrm{leads}} = \sum_{n=1}^{N} \frac{v_s}{2\pi}\int_{0}^{L}dx[K(\partial_{x}\theta_{n})^{2}
+ K^{-1}(\partial_{x}\phi_{n})^{2}]$, where the density fluctuation modes of the TLLs are given by the operator $\rho_{n}(x)\sim \partial_{x}\phi_{n}(x)/\pi$ and the canonically conjugate momenta by 
$\partial_{x}\theta_{i}(x)$.
Direct hopping processes between the leads can be described using
the boson annhilation/creation operators
$\psi_{n}(x)\sim e^{i\theta_{n}(x)}\vert_{x=0}$
at the end-points of the semi-infinite TLLs:
$H_{\mathrm{tunn}} = \thopp\,\sum\nolimits_{n,n^{\prime}}
(e^{i(\theta_{n}(0)-\theta_{n^{\prime}}(0))}+
{\rm h.c.}). $
At low frequencies, the ground state of the complete system consists of either
$N$ uncorrelated wires
(i.e., the disconnected fixed point, DFP, $\thopp =0$), or $N$ maximally correlated wires (i.e., the connected fixed point, CFP, $\thopp\to\infty$).
The CFP manifests itself via the mutual pinning of the phase fields at the junction: $\theta_{n}(x,t)=
\theta_{n^{\prime}}(x,t)\vert_{x=0}$ for all pairs $(n,n^{\prime})$~\cite{tokuno}.
Additionally, current conservation demands that $\sum_{n=1}^{N}\partial_{t}\theta_{n}(0,t)=0$.
In the language of the renormalization group (RG), the hopping $\thopp$ flows to $0$
(i.e., DFP) for $K<1$~\cite{kane}, and to strong coupling (CFP) for
$K>1$~\cite{kane,tokuno}.
An RG analysis around the CFP for the
backscattering amplitude $\lambda$ gives ${d\lambda}/{dl} =
\left\{1-\left(2(N-1)K/N\right)\right\}\lambda$.  Therefore, the CFP is stable for
$K> N/\left(2(N - 1)\right)$.
Both the DFP and the CFP are stable for $N/[2(N-1)]<K<1$;
for $K$ in this interval there must
therefore be a quantum phase transition at some $\thopp^* \neq 0$, separating the ground states of $N$ uncorrelated and $N$ maximally
correlated wires.  This transition is analogous to that exhibited by a quantum Brownian particle on an
$(N-1)$-dimensional triangular lattice
in the presence of Ohmic dissipation~\cite{yikane}.
\begin{figure}[htb]
\begin{center}
\scalebox{1.0}{
\includegraphics{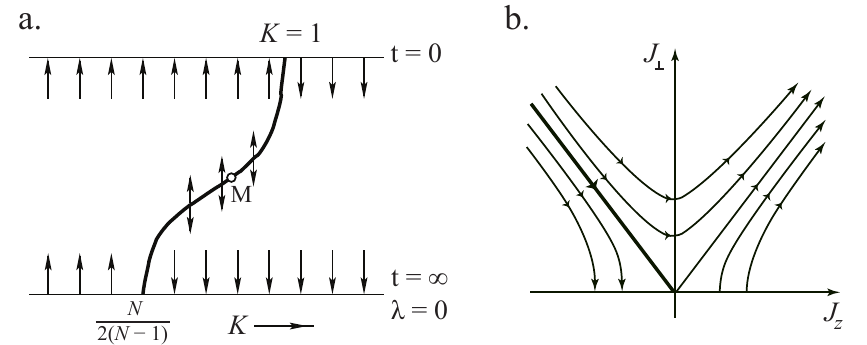}}
\end{center}
\vspace*{-0.6cm}
\caption{
(a)~Phase diagram for $N$ leads with interlead hopping $\thopp$ and TLL parameter $K$.
The phase boundary (thick line) separates the
disconnected (i.e., DFP, $\thopp =0$) and maximally
connected (i.e., CFP, $\thopp\rightarrow \infty$) fixed points, and has a self-dual point (\textbf{M}).
(b)~Kosterlitz-Thouless flow for lead-dot couplings at fixed $(K, \thopp)$.
The left separatrix (thick line) demarcates the boundary between the ferromagnetic and Kondo phases.}
\label{scattrg}
\end{figure}
\par
The novel feature of the $N$-wire phase diagram shown in Fig.~\ref{scattrg}a is that the CFP is stable against weak backscattering close to the junction, even for $K<1$ (in contrast with the two-wire case~\cite{kane}).
This occurs because
the boundary conditions on the fields at the junction imply that a wavepacket
arriving at the junction from any one wire
meets an effective composite TLL,
comprising the $N-1$ other leads, for which  $K_{\mathrm{eff}} = (N - 1) K > 1$.
The locking of the phase field in the first wire to that of the composite TLL precludes any
chemical-potential drop across the junction, although only a fraction of the incoming current enters any individual lead of the composite TLL~\cite{tokuno}.  Said another way, the local inertia of the phase fields strongly suppresses backscattering events involving high-momentum phase fluctuations.
This phenomenon is dual to the enhanced inertia in the
number fields at the endpoints of the wires at the DFP for $K<1$, which results in a power-law suppression of the tunneling density of states (TDOS).
\par
We now incorporate the dot by coupling it to the $N$ leads via $H^n_{\mathrm{int}}$ [see Eq.~(\ref{eq:irlm})], and develop the phase diagram via an RG analysis around
the CFP and DFP. Near the CFP, we find the following scaling equations:
\beq
\label{eq:KTone}
\frac{dJ_{z}}{dl} = J_{\perp}^{2}\,,\quad
\frac{dJ_{\perp}}{dl} =
\left(1-\frac{2(N-1)K}{N} + J_z \right)J_{\perp}\,,
\eeq
to second order in all couplings, where the couplings have been scaled
by the high-frequency cutoff $\omega_{c}=v_s/\xi$ (where $\xi$ is the healing length~\cite{recati}). In addition, we find that $d\lambda/dl = \left\{1-\left(2(N-1)K/N\right)\right\}\lambda + (J^2/\omega_c)$.
By shifting $J_{z}$ to
$J'_{z}= J_{z}+1-2(N-1)K/N$,
Eqs.~(\ref{eq:KTone}) assume the well-known
Kosterlitz-Thouless form~\cite{kosthou}
(see Fig.~\ref{scattrg}b),
with a Kondo temperature scale given by
$T_{K}\sim \omega_c e^{-1/J'_{z}}$.
Thus, $J_{z}$ is found to be RG-marginal at first order but RG-relevant at second order, and independent of $K$ and $N$.
On the other hand, $J_{\perp}$ has the same scaling dimension as
$\lambda$, and is thus dependent on $K$ and $N$.
For $K>\left[N/2(N-1)\right]$,
even though $J_{\perp}$ is RG-irrelevant
(from its scaling dimension),
it can turn relevant, due to the growth of $J_{z}$.
The scaling equation for $J_{\perp}$ admits a nontrivial
fixed point at
$\tilde{J}_{z}\equiv
\left[2(N-1)K/N\right]-1$.
For $J_{z}>\tilde{J}_{z}$, all flows lead to the $N$-channel Kondo fixed point;
for $J_{z}<\tilde{J}_{z}$, flows lead to zero dot-to-lead hopping.
\par
If $K <\left[N/2(N-1)\right]$,
the RG analysis about the CFP is invalid;
one must instead analyze the Kondo couplings around the DFP.
The scaling relations near the DFP are
\beq
\label{eq:KTtwo}
\frac{dJ_{z}}{dl} = J_{\perp}^{2}\,,\quad \frac{dJ_{\perp}}{dl} =
\left(1-\frac{1}{K} + J_z \right)J_{\perp} \,.
\eeq
As with $\lambda$, the flow for $\thopp$ acquires a positive contribution of order $J_\perp^2/\omega_c$ from the dot-mediated hopping (i.e., the dot promotes interlead hopping, as one might expect). By shifting $J_{z}$ to
$J'_{z}= J_{z}+1-(1/K)$, Eq.~(\ref{eq:KTtwo}) assumes the Kosterlitz-Thouless form.
Similarly, for $K<1$ the scaling equation for $J_{\perp}$ has a nontrivial fixed point at $J_{z}^{*}=(1-K)/K$.
For $J_{z}>J_{z}^{\ast}$, all flows lead to the $N$-channel Kondo fixed point;
for $J_{z}<J_{z}^{\ast}$, flows lead to zero dot-to-lead hopping.
\begin{figure}[htb]
\begin{center}
\scalebox{1.0}{
\includegraphics{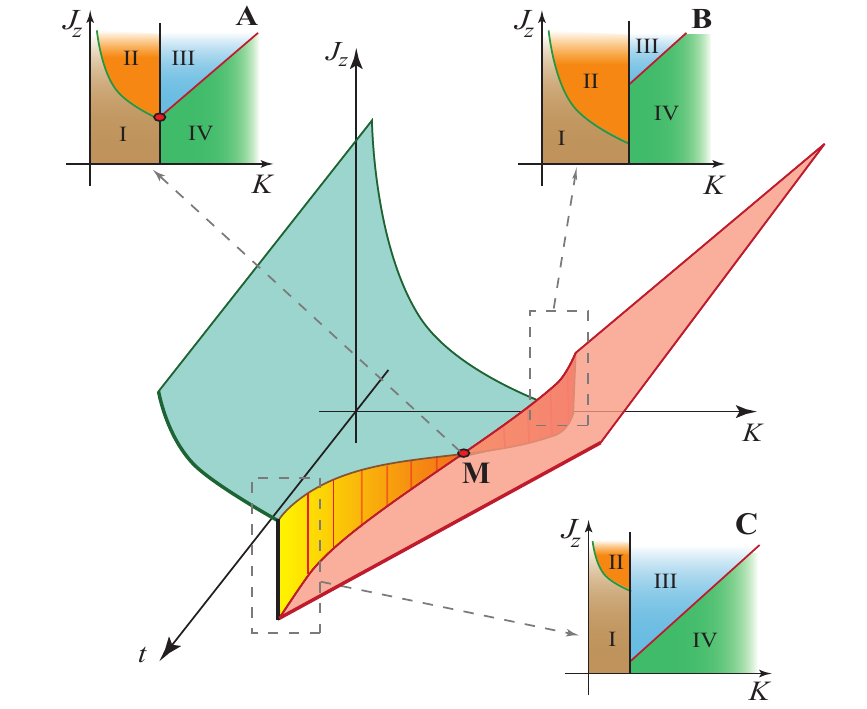}}
\end{center}
\vspace*{-0.6cm}
\caption{Phase diagram of the dot-lead system.
The discontinuous critical surface separates the Kondo and ferromagnetic phases.
The curved vertical ribbon (orange) is the phase boundary
between the DFP and CFP phases
(see Fig.~\ref{scattrg}).
The discontinuity of the critical surface shrinks to zero at the multicritical point~\textbf{M}.
The diagram has four phases---
I: completely decoupled wires;
II: wires coupled only through dot;
III: wires connected both directly and through dot;
IV: wires connected but decoupled from dot.
The transitions between them (shown in insets~A, B, and C)
are described in the text.}
\label{kondorg}
\end{figure}
\par
Bringing together the flows of
$(\thopp,J_{\perp},J_{z})$
yields the three-dimensional phase diagram shown in Fig.~\ref{kondorg}.
The tuning of $\thopp$ and/or $K$ allows one to access two nontrivial transitions between phases that
have opposing characters in {\it both\/} their Kondo
coupling to the dot and their direct interlead hopping.
One is a transition between phase~II
(in which $N$-channel Kondo physics dominates TDOS suppression)
and phase~IV
(in which Kondo physics is suppressed by local phase pinning); see Fig.~\ref{kondorg}B.
The other is a transition between phase~I
(in which Kondo physics is dominated by TDOS suppression)
and phase~III
(in which Kondo screening overcomes local phase pinning);
see Fig.~\ref{kondorg}C.
In addition, we find a multicritical point
(see point~M in Fig.~\ref{kondorg}A)
at intermediate coupling in $\thopp$;
this occurs when $\tilde{J}_{z}=J^{*}_{z}$ [so that $K=\sqrt{N/2(N-1)}$].
Point~M coincides with the self-dual point~\cite{yikane} in the phase
boundary of intermediate-coupling fixed points (see Fig.~\ref{scattrg}a),
and involves a compromise between the competing tendencies
of TDOS suppression and local phase pinning.
For the special case of $N=2$, the phase boundary in Fig.~\ref{kondorg}a becomes
a marginal line at $K=1$ (i.e., the Tonks-Girardeau gas~\cite{paredes}),
and the point~M becomes a multicritical line at $\tilde{J}_{z}=J^{*}_{z}=0$.
\par
The four phases are characterized by the following bare two-lead transmission coefficients
across the junction: $G_{2}=0$ (i.e., minimal) at the DFP and
$G_{2}=4K/N^{2}$ (i.e., maximal) at the CFP.
These bare coefficients acquire power-law corrections of order
$\thopp^{2}\,T^{\nu}$ (DFP) and $\lambda^{2}\,T^{\mu}$ (CFP),
arising from direct interlead scattering~\cite{kane};
here, $T$ represents an energy scale (e.g., temperature) that cuts off the RG flows,
and $(\nu,\mu)$ are exponents determined by the leading irrelevant perturbations
around the corresponding fixed point.
In experiments with cold atoms, these power-law contributions should be detectable
via real-time lead dynamics~\cite{tokuno}.
\begin{figure}
	\centering
		\includegraphics{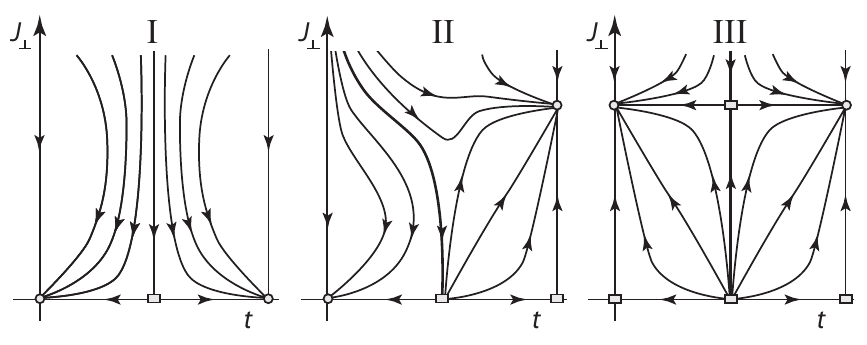}
\vspace*{-0.6cm}
\caption{RG flows in the $t$-$J_\perp$ plane for $N/2(N-1) < K <\sqrt{N/2(N-1)}$.
Here, $J_\perp \ll |J_z - J_z^*|, |J_z - \tilde{J}_z|$.
I.~$J_z < J_z^*, \tilde{J_z}$:
hopping via the dot is irrelevant, and the dot stays disconnected. II.~$\tilde{J}_z < J_z < J_z^*$: for $J_\perp \rightarrow 0$, hopping via the dot is irrelevant for $\thopp < \thopp^*$ and relevant otherwise. However, sufficiently large $J_\perp$ can drive $\thopp$ past $\thopp^*$ toward the regime in which both $\thopp$ and $J_\perp$ grow at low energies.
III.~$J_z > J_z^*$:
hopping via the dot is relevant on both sides.
Figures~I and III would remain identical for
$\sqrt{N/2(N-1)} < K <1$,
while II would become inverted about the dashed line.}
	\label{fig:rgflow2}
\end{figure}
\par
Number fluctuations on the dot can be accessed via a non-destructive measurement
scheme such as, e.g., that suggested in Ref.~\cite{ritsch07} for the Bose-Hubbard model: in such a scheme the dot
would be located in the waist of a high-finesse optical cavity that has a resonance
frequency near an optical transition of the hyperfine state $b$ (but not of $a$).
A fixed-number state on the dot (i.e., the unscreened spin) would merely shift the
cavity's resonance; by contrast, a fluctuating-number state (i.e., the Kondo state) would
lead to a double-peak structure in the transmission spectrum of the cavity, with peaks
corresponding to an empty dot and to an occupied dot.
\noindent\textit{Acknowledgments}.
We thank E.~Demler and M.~Pasienski for stimulating discussions.
This work was supported by DOE Award No.~DE-FG02-07ER46453 (S.L., P.M.G.),
and NSF DMR-0605813 (S.L.) and DMR~09-06780 (S.G.)

\end{document}